\documentclass[twocolumn,showpacs,preprintnumbers]{revtex4}%
\usepackage{amsmath}
\usepackage{amsfonts}
\usepackage{amssymb}
\usepackage{graphicx}%
\providecommand{\U}[1]{\protect\rule{.1in}{.1in}}
\begin{document}
\title{Minimal Multifractality in the Spectrum of a Quasiperiodic Hamiltonian}
\author{Gerardo G. Naumis}
\affiliation{Instituto de F\'{\i}sica, Universidad Nacional Aut\'{o}noma de M\'{e}xico
(UNAM), Apartado Postal 20-364, 01000, M\'{e}xico, Distrito Federal, Mexico.}
\date{\today }

\begin{abstract}
In many systems, the electronic energy spectrum is a continuous or singular
continuous multifractal set with a distribution of scaling exponents. Here, we
show that for a quasiperiodic potential, the multifractal energy spectrum can
have a minimal dispersion of the scaling exponents. This is made by tuning the
ratio between self-energies in a tigth-binding Hamiltonian defined in a
quasiperiodic Fibonacci chain. The tuning is calculated numerically from a
trace map, and coincides with the place where the scaling exponents of the
map, obtained from cycles with period six and two, are equal. The diffusion of
electronic wave packets also reflects the minimal multifractal fractal nature
of the spectrum, by reducing intermittency. The present result can help to
simplify the task of studying several problems in quasiperiodic systems, since
the effects of multiscaling are better isolated.

\end{abstract}

\maketitle
\affiliation{$^{1}$Instituto de F\'\i sica, Universidad Nacional Aut\'onoma}
\affiliation{de M\'exico (UNAM)}
\affiliation{Apartado Postal 20-364, 01000, Distrito Federal, M\'exico.}

Quasiperiodic systems, which are neither periodic, nor disordered, have been
the focus of an active research since the seventies \cite{Hofstadter,Aubry}.
Later on, they gained much more importance when the discovery of alloys with
fivefold symmetry and long range order was made \cite{Blech}. These alloys
were called quasicrystals, in recognition of their quasiperiodic atomic
structure. The word quasiperiodic means that the structure can be represented
as a sum of periodic functions, each with an incommensurate period with
respect to the others. Still, the nature of the physical properties of
quasicrystals is not well understood \cite{Thiel}\cite{Vedemenko}\cite{Moras}.
Even in the theoretical side there is a lack of understanding in how electrons
propagate, specially in two and three dimensions \cite{Maciarep}. As is well
known, a periodic potential satisfies the Bloch%
%TCIMACRO{\U{b4}}%
%BeginExpansion
\'{}%
%EndExpansion
s theorem, which tells that the eigenstates of the Schr\"{o}edinger equation
are plane waves of delocalized nature, and the energy spectrum is continuous
\cite{Kittel}. For disordered systems, like in the one dimensional (1D)
Anderson model, all the states are localized corresponding to isolated
eigenvalues \cite{Zallen}. In more dimensions, there is a mobility edge which
separates extended from localized states \cite{Zallen}. For most of the
quasiperiodic systems in 1D, the spectrum is neither continuous nor singular,
instead a new type of spectrum, called singular continuous is obtained
\cite{Suto}. This kind of spectrum is similar to a Cantor set, and presents a
multifractal nature. The corresponding eigenfunctions are called critical, and
also show self-similarity and multifractality. In two and three dimensions,
the nature of the spectrum is not known, although there seems to be a kind of
mobility edge \cite{NaumisPenrose}\cite{NaumisJPC}. However, even in 1D, where
large amount of work has been done, there are many unsolved questions, like
the nature of conductivity \cite{Dominguez} or diffusivity \cite{Ketmerick},
the spectral statistics and the shape of many of the eigenfunctions
\cite{Fujiwara}\cite{Macia}. \ For example, Machida \textit{et. al.}
\cite{Machida} showed that for certain parameters of the quasiperiodic Harper
equation, the distribution of level spacing follows an inverse power law. This
tendency was explained by Geisel et. al. \cite{Geisel} as a level clustering.
More recently \cite{Evangelou}, it has been argued that the clustering regimen
of the distribution of level spacings in quasiperiodic potentials was an
artifact of an inappropriate way of making the unfolding procedure, and the
fact that the distribution of gaps and bands were different was considered as
a signature of different scaling indices in the spectrum \cite{Evangelou}.

Many of the difficulties that arise in the studies of quasiperiodicity, are
due to the fact that all of the systems that have been used up to now, present
a multifractal spectrum \cite{Ketmerick}\cite{Geisel}\cite{Evangelou}%
\cite{Tang2}, \textit{i.e.}, the scaling with the system size is not given by
only one scaling exponent ($\alpha$), instead a distribution of exponents
determined by a function called $f(\alpha)$ is usually found. This makes very
difficult to relate the spectrum with the corresponding eigenfunctions, since
they have a complicate shape due to \textit{multiscaling }and
\textit{intermittency}. For example, the moments of a wave packet that is
diffusing behave as $<x^{q}(t)>\sim t^{q\beta(q)}$ for a time $t$, where
$\beta(q)$ is an exponent related with the spectral type. \ In principle, it
seems that one needs to exclude any simple relation between spectral and
diffusion exponents \cite{Ketmerick}\cite{Piechon}. However, as we show in
this letter, the problem can be more transparent if the parameters of the
quasiperiodic potential are chosen in such a way that the system presents a
minimal dispersion of scaling exponents. As a consequence, instead of dealing
with a wide multifractal set, which makes all computations very demanding from
a technical point of view, one considers a very narrow multifractal set, which
is much closer to a pure monofractal. Thus, it is possible to improve the
isolation of \textit{intermittency, }without dealing with huge
\textit{multiscaling.}

As starting point, we use the simplest model of a quasicrystal: the Fibonacci
chain (FC). The FC is build as follows: consider two letters, $A$ and $B$, and
the substitution rules, $A\rightarrow B,$ and $B\rightarrow AB.$ If one
defines the first generation sequence as $\mathcal{F}_{1}=A$ and the second
one as $\mathcal{F}_{2}=BA$, the subsequent chains are generated using the two
previous rules, for instance, $\mathcal{F}_{3}=ABA.$ Starting with an $A$, we
construct the following sequences, $A,$ $B$,$AB$, $BAB,$ $ABBAB,$ $BABABBAB,$
and so on. Each generation obtained by iteration of the rules is labeled with
an index $l.$ Is clear that the number of letters in each generation $l$ is
given by the Fibonacci numbers $F(l)$ of generation $l $, which satisfy:
$F(l)=F(l-1)+F(l-2)$ with the initial conditions: $F(0)=1,F(1)=1.$ A natural
model for a one dimensional quasicrystal is a tight-binding hamiltionian of
the type,%

\begin{equation}
E\psi_{n}=V(n)\psi_{n}+t_{n}\psi_{n+1}+t_{n-1}\psi_{n-1} \label{tb}%
\end{equation}
where $\psi_{n}$ is the wave-function at site $n$, $t_{n}$ is the resonance
integral between sites $n$ and $n+1$. $V(n)$ is the atomic on-site potential
and $\ E$ are the allowed energies. For the present purposes, $t_{n}$ is set
to $1$ for all sites, and $V(n)$ has two possible values, $V_{A}$ and $V_{B}$
following the Fibonacci sequence. \ In fact, using a Fourier expansion, we can
show that the Fibonacci potential is no more than a sum of well known Harper
potentials,%
\begin{equation}
V(n)=\overline{V}+2\delta V%
%TCIMACRO{\dsum \limits_{s=1}^{\infty}}%
%BeginExpansion
{\displaystyle\sum\limits_{s=1}^{\infty}}
%EndExpansion
\widetilde{V}(s)\cos\left(  \pi s\phi(2n+1)\right)  \label{pot}%
\end{equation}
where $\phi$ is the golden mean $\phi=(\sqrt{5}+1)/2$, $\overline{V}%
=(V_{A}/\phi)+(V_{A}/\phi^{2}),$ and $\widetilde{V}(s)$ is the $s$ harmonic of
the Fourier series, $\widetilde{V}(s)=\sin(\pi s\phi)/s\phi.$ The strength of
the quasiperiodicity is measured by the difference between site-energies
$\delta V=V_{A}-V_{B}$.

The spectrum of this problem, \textit{i.e.}, the energies that satisfy
eq.(\ref{tb}), are obtained by using the trace of the transfer matrices. These
energies are those for which the trace of generation $l$ (that we denote by
$2x_{l}(E)$) of the transfer matrix satisfy $-1\leq x_{l}(E)\leq1$. In the
case of a FC, $x_{l}(E)$ can be obtained from a non-linear map given by
\cite{Kadanoff},
\begin{equation}
x_{l+1}(E)=2x_{l}(E)x_{l-1}(E)-x_{l-2}(E), \label{map}%
\end{equation}
with the initial conditions $x_{-1}=1,x_{0}=(E+\lambda)/2,x_{1}=(E-\lambda
)/2,$ and $\lambda=\left\vert \delta V\right\vert /2$. Since the map contains
three successive generations, one can consider the evolution of the trace as a
trajectory in a 3D space. It has also been shown that the spectrum is a
multifractal set \cite{Tang}, and it was conjectured that the exponents
describing the scaling of the spectrum near the upper band edge and band
center were governed by cycles of periods $6$ and $2$ of the trace map. Using
a linearized stability analysis around these cycles, two scaling indices were
found \cite{Oono}, one corresponding to the period $6$ cycle,
\begin{equation}
\alpha_{6}=\frac{\ln\left\{  \left[  1+4(1+\lambda^{2})^{2}\right]
^{1/2}+2(1+\lambda^{2})\right\}  ^{2}}{\ln\phi^{6}}, \label{ac6}%
\end{equation}
and the other to the period $2$ cycle,
\begin{equation}
\alpha_{2}=\frac{\ln\left\{  8J-1+\left[  (8J-1)^{2}-4\right]  ^{1/2}\right\}
/2}{\ln\phi^{2}} \label{a2}%
\end{equation}
with $J=[3+(25+16\lambda^{2})^{1/2}]/8$. In Ref. \cite{Tang} , it was
concluded that $\alpha_{2\text{ }}$ was the minimum scaling exponent, related
with the band-width of the upper band edges, $\alpha_{6\text{ }}$ was the
maximal scaling exponent. Between them, a continuous distribution of exponents
was found. That argument can be related with the shape of the wave-function,
and in principle it is possible to conclude that wave-functions in the edges
decay more rapidly than in the center of the spectrum \cite{Tang}. Later on,
it was proved that the $6$ cycle does not always occur at the center or edges
of the spectrum, and in fact, the question of which is the maximal contraction
factor is an open question, since there are other cycles \cite{Rudinger}. In
spite of this, in fig. 1 we show a plot of scaling exponents from cycles $6$
and $2$, where is clearly seen that not always $\alpha_{2\text{ }}%
<\alpha_{6\text{ }}$, as was erroneously stated in ref. \cite{Tang}.
%TCIMACRO{\FRAME{ftbpFU}{3.9591in}{2.783in}{0pt}{\Qcb{Scaling exponents
%$\alpha_{6}$ (solid line) and $\alpha_{2}$ (dotted line) as a function of
%$\delta V$. The arrow indicates the place where $\alpha_{6}=\alpha_{2}.$ The
%numerical results are shown as circles. Inset: absolute value of the
%difference between $\alpha_{\max}-\alpha_{\min}$ for diferent values of
%$\delta V.$ The triangules, squares and circles are the numerical results for
%generations $l=9,12$ and $15$ respectively$.$ The solid line is the absolute
%value of the difference between $\alpha_{6}$ and $\alpha_{2}.$}}%
%{\Qlb{scaling}}{scalingexpo.eps}{\special{ language "Scientific Word";
%type "GRAPHIC";  maintain-aspect-ratio TRUE;  display "USEDEF";
%valid_file "F";  width 3.9591in;  height 2.783in;  depth 0pt;
%original-width 8.489in;  original-height 5.9525in;  cropleft "0";
%croptop "1";  cropright "1";  cropbottom "0";
%filename 'scalingexpo.eps';file-properties "XNPEU";}}}%
%BeginExpansion
\begin{figure}
[ptb]
\begin{center}
\includegraphics[
height=2.783in,
width=3.9591in
]%
{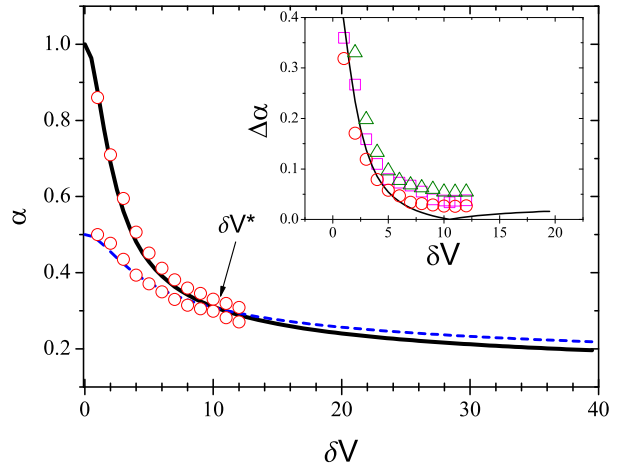}%
\caption{Scaling exponents $\alpha_{6}$ (solid line) and $\alpha_{2}$ (dotted
line) as a function of $\delta V$. The arrow indicates the place where
$\alpha_{6}=\alpha_{2}.$ The numerical results are shown as circles. Inset:
absolute value of the difference between $\alpha_{\max}-\alpha_{\min}$ for
diferent values of $\delta V.$ The triangules, squares and circles are the
numerical results for generations $l=9,12$ and $15$ respectively$.$ The solid
line is the absolute value of the difference between $\alpha_{6}$ and
$\alpha_{2}.$}%
\label{scaling}%
\end{center}
\end{figure}
%EndExpansion

Furthermore, there is a cross-over of the two scaling exponents. The exponent
$\alpha_{2}$ is the minimum only from $\delta V=0$ to $\delta V\approx10.5.$
The crossover occurs when $\alpha_{2}=\alpha_{6}$. The special value of
$\delta V^{\ast}$ where this happens can be calculated by making Eq.
(\ref{ac6}) equal to Eq. (\ref{a2}). After many lengthy algebraic steps, it is
possible to show that $\delta V^{\ast}$ is given by the irrational number,%
\[
\delta V^{\ast}=2\sqrt{13+\frac{11}{2}\sqrt{7}}\approx
10.497929740830855719340651...
\]
This value perfectly agrees with the plot of $\alpha_{6}$ and $\alpha_{2}$. In
the special value $\delta V^{\ast},$ $\alpha_{6}=\alpha_{2}=d_{f}^{\ast
}\approx0.305$. If as stated in ref. \cite{Tang}, the $6$ and $2$ cycles were
the extremal scaling exponents, then $\alpha_{\max}=$ $\alpha_{\min}$ at
$\delta V^{\ast},$ and thus one should expect only one scaling exponent,
corresponding to a pure monofractal. To test how far is this assumption from
the real behavior of the spectrum, we have performed numerical calculations.
First, the band widths of the spectrum were obtained by the iteration of the
map (\ref{map}), using quadruple precision ($32$ significant digits) for each
value of $\delta V.$ The calculations were made for chains of generations
$l=9,12$ and $15$. The maximal and minimal scaling exponents have the
following expression,%
\begin{equation}
\alpha_{\max}=\frac{\ln\Delta E_{0}-\ln F(l)}{\ln\Delta E_{\min}(l)}%
,\alpha_{\min}=\frac{\ln\Delta E_{0}-\ln F(l)}{\ln\Delta E_{\max}(l)}%
\end{equation}
where $\Delta E_{0}$ is the initial band width, and $\Delta E_{\min}(l)$
($\Delta E_{\max}(l)$) are the minimal (maximal) bandwidths of a $l$
generation FC. The contribution of $\Delta E_{0}$ tends to zero in an infinite
system, but is needed in the present finite case. Figure 1 shows the numerical
values of $\alpha_{\max}$ and $\alpha_{\min}$ for $l=15$ compared with the
theoretical values. In spite that $\alpha_{6}$ and $\alpha_{2}$ are not the
maximal and minimal scaling exponents, they are close to the numerical values,
and a minimal dispersion of the spectrum is seen at $\delta V^{\ast}$. The
reason of the very close values of the real extremal exponents, $\alpha_{\max
}$ and $\alpha_{\min},$ compared with $\alpha_{6}$ and $\alpha_{2}$, is
because these cycles dominates the spectrum when $\delta V\rightarrow\infty.$
Notice that beyond $\delta V=12,$ there are no numerical points due to the
increased difficulty in\ computing the band widths, since a lot of precision
is required as $\delta V\rightarrow\infty.$

It is also possible to calculate the multifractal distribution $f(\alpha)$ of
scaling exponents. Let us denote the bandwidth of the $i-$th band of the
spectrum by $\Delta E_{i}$. Each band contains the same measure or density of
states $1/F(l)$. Each band scales as: ($\Delta E_{i})^{\alpha}=1/F(l).$ A
partition function is defined by \cite{Halsey},
\begin{equation}
\Gamma_{l}(q,\tau)=\sum_{i=1}^{F(l)}\left(  \frac{1}{F(l)}\right)  ^{q}\left(
\frac{1}{\Delta E_{i}}\right)  ^{\tau}%
\end{equation}
This function has a limit when $l$ goes to infinity, which is either zero or
infinity, unless $\tau$ and $q$ are chosen in an appropriate way \cite{Halsey}
such that $\Gamma_{l}(q,\tau)=1.$ This condition determines a function
$\tau(q)$. The fractal dimension for a set of points with the scaling $\alpha
$, is obtained by a Legendre transformation $f(\alpha)=-\tau(q)+\alpha q$,
where $\alpha=d\tau(q)/dq$. In figure 2, the evolution of $f(\alpha)$ versus
$\alpha$ is shown for different values of $\delta V$ using generation $l=9$ .
Notice how the width of the curves decrease and in fact is a minimum as
$\delta V\rightarrow\delta V^{\ast}.$ The width of the curve $f(\alpha)=0$
also defines the values of $\alpha_{\min}$ and $\alpha_{\max}$. This can also
bee seen in \ the inset of fig. 1, where we plot the absolute value of the
difference $\Delta\alpha=\left\vert \alpha_{\max}-\alpha_{\min}\right\vert $
for different generations. Observe that $f(\alpha)$ has a finite width at
$\ \delta V^{\ast}$, and thus $\ \alpha_{\min}\neq\alpha_{\max}.$ This rules
out the possibility of having a pure monofractal spectrum, which was something
to be desired. However, the numerical calculations suggests that $\delta
V^{\ast}$ is the value where the spectrum has a minimal multifractal dispersion.%

%TCIMACRO{\FRAME{ftbpFU}{3.4662in}{2.437in}{0pt}{\Qcb{Evolution of $f(\alpha)$
%as $\delta V$ is changed for a FC of generation $l=9$. From right to left,
%$\delta V=3,4,5,6,7,8,9,\delta V^{\ast},11,12.$ The special value $\delta
%V^{\ast}$ is indicated with an arrow. As an example, $\alpha_{\max}$ and
%$\alpha_{\min}$ are indicated for $\delta V=3.$}}{\Qlb{falfas}}{falfas.eps}%
%{\special{ language "Scientific Word";  type "GRAPHIC";
%maintain-aspect-ratio TRUE;  display "USEDEF";  valid_file "F";
%width 3.4662in;  height 2.437in;  depth 0pt;  original-width 8.489in;
%original-height 5.9525in;  cropleft "0";  croptop "1";  cropright "1";
%cropbottom "0";  filename 'falfas.eps';file-properties "XNPEU";}}}%
%BeginExpansion
\begin{figure}
[ptb]
\begin{center}
\includegraphics[
height=2.437in,
width=3.4662in
]%
{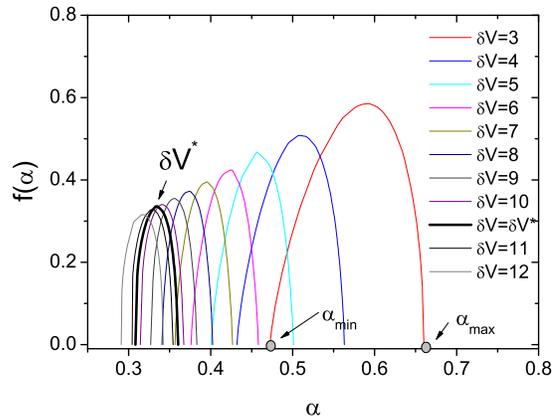}%
\caption{Evolution of $f(\alpha)$ as $\delta V$ is changed for a FC of
generation $l=9$. From right to left, $\delta V=3,4,5,6,7,8,9,\delta V^{\ast
},11,12.$ The special value $\delta V^{\ast}$ is indicated with an arrow. As
an example, $\alpha_{\max}$ and $\alpha_{\min}$ are indicated for $\delta
V=3.$}%
\label{falfas}%
\end{center}
\end{figure}
%EndExpansion

Since there is a close relationship between band scaling, localization and the
dispersion of electronic wave functions \cite{Naumis}, one can expected to
observe the signature of minimal multifractality in the electron diffusion. A
way of characterizing the propagation consists in the study of the
wavefunction moments as a function of time \cite{Ketmerick}, which are defined
as,
\[
\left\langle x^{q}(t)\right\rangle =\sum_{n=1}^{F(l)}(n-n_{0})^{q}\Psi^{\ast
}(n,t)\Psi(n,t)
\]
where $j$ is the site position in the chain. $\Psi(n,t)$ is the wavefunction
at time $t$ on site $n,$ that evolves following the time dependent
Schr\"{o}edinger equation, in which $E$ is replaced by the operator
$-i\partial/\partial t$ . As was already stated, in quasiperiodic systems
$\left\langle x^{q}(t)\right\rangle $ follows a power law behavior of the type
$\left\langle x^{q}(t)\right\rangle =\mathcal{D}(q)t^{q\beta(q)},$ where
$\mathcal{D}(q)$ is a constant, known as the diffusion constant for the
periodic case. In the present work, we have performed numerical simulations to
obtain $\beta(q)$ by using chains of generation $l=15$. As initial conditions,
a delta function wave-packet is drop at time $t=0$ at site $n_{0}$. Different
initial sites were used, but the results are almost $n_{0}$ independent. The
numerical computations were made by solving the time dependent
Schr\"{o}edinger equation by two methods that gave similar results. The first
was a direct numerical resolution of the differential Schr\"{o}edinger
equation using a discretization of time, and the second consists in a
decomposition in eigenfunctions. Since each eigenfunction evolves
independently with time, the resulting wave function at any time is just a
linear combination of eigenfunctions, where the coefficients of the
combination are the projections of the initial condition into each
eigenvector. For long times, the second method is much more accurate. Finally,
the scaling exponents $\beta(q)$ were obtained by fitting a power law in a
log-log plot of $\left\langle x^{q}(t)\right\rangle $ versus $t$. In all of
the cases, the fit was made from $t=1000$ to $t=50000$ (all units have been
set to one) in order to avoid initial transients.

\bigskip%
%TCIMACRO{\FRAME{ftbpFU}{3.6538in}{2.5694in}{0pt}{\Qcb{$\beta(q)$ as a function
%of the difference in self-energies $\delta V$ for $q=2,4,6 $ and $8$. A
%rigorous bound for $\beta(q)$ is shown with a grey line. Also, the scaling
%exponents $\alpha_{6}$ and $\alpha_{2}$ are shown with solid and dotted lines
%respectively.}}{\Qlb{moments}}{moments.eps}%
%{\special{ language "Scientific Word";  type "GRAPHIC";
%maintain-aspect-ratio TRUE;  display "USEDEF";  valid_file "F";
%width 3.6538in;  height 2.5694in;  depth 0pt;  original-width 8.489in;
%original-height 5.9525in;  cropleft "0";  croptop "1";  cropright "1";
%cropbottom "0";  filename 'moments.eps';file-properties "XNPEU";}}}%
%BeginExpansion
\begin{figure}
[ptb]
\begin{center}
\includegraphics[
height=2.5694in,
width=3.6538in
]%
{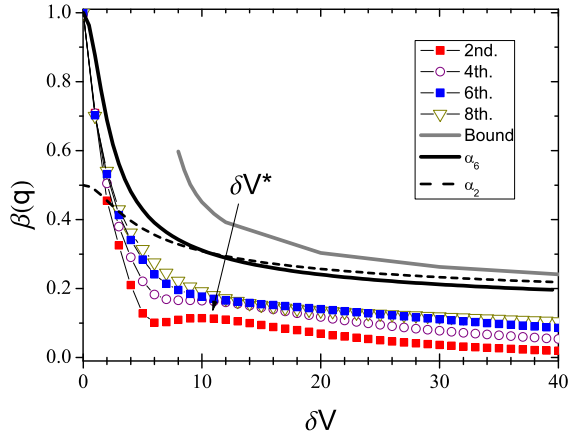}%
\caption{$\beta(q)$ as a function of the difference in self-energies $\delta
V$ for $q=2,4,6 $ and $8$. A rigorous bound for $\beta(q)$ is shown with a
grey line. Also, the scaling exponents $\alpha_{6}$ and $\alpha_{2}$ are shown
with solid and dotted lines respectively.}%
\label{moments}%
\end{center}
\end{figure}
%EndExpansion

Figure 3 shows the values of $\beta(q)$ as a function of $\delta V$ for the
first eight moments. Observe that odd moments are not displayed since they are
zero. The graph shows a very interesting feature which is a nearly collapse of
$\beta(q)$ for $q=4,6$ and $8$ at $\delta V^{\ast}$. This suggests that
$\beta(q)\rightarrow\beta^{\ast}\approx0.16$ for such values of $q.$ A local
maximum is obtained for $q=2$. Since $\beta^{\ast}$ is nearly $q$ independent,
and $\beta(q)$ is different from one, is clear that the wave functions show
intermittency\textit{\ }but almost no multiscaling, although this effect is
mitigated since $\beta(2)$ has a lower value. In figure 3 we have also
included the values of $\ \alpha_{\min}$ and $\alpha_{\max}$ as a comparison.
Notice the crossover near the collapse of $\beta(q).$ The fact that $\beta(q)$
and $\alpha$ are represented in the same graph is not just a coincidence,
since the scaling of the spectrum is related with wave-localization. In fact,
the present work suggests that in a very rough approximation $d_{f}^{\ast
}\approx\beta^{\ast}/2$, which is similar to the result obtained from a
analysis of the relationship between wave localization overlap and the
spectrum \cite{NaumisJPCM}. Also, in figure 3 we present a comparison with a
strict mathematical bound that has recently been proved \cite{Damanik} for
$\beta(q)$ when $\delta V>8.0.$ The bound works as predicted, although is
higher than the actual values of $\beta(q).$

In conclusion, we have shown a value of the Fibonacci potential for which
there is a minimal dispersion in the scaling exponents of the multifractal
spectrum. The value of the potential seems to de determined mainly by the six
and two cycles. The behavior of wave packet diffusion reflects this fact by
showing almost no multiscaling. We hope that the present work will bring much
more understanding into the properties of quasiperiodic Hamiltonians, since
one can better isolate effects of multiscaling. For example, it would be
useful to study the evolution of the spectral statistics near this minimal
condition. Since the trace map is similar for phonons or the case of
quasiperiodic hopping integrals, the presented property can be extended to
such cases, or different trace maps \cite{Grimm}.

I would like to thank DGAPA-UNAM project IN-117806 for financial help.

\end{document}